\documentclass[10pt]{IEEEtran}

\usepackage[cmex10]{amsmath}
\usepackage{amsfonts}
\usepackage[font=footnotesize]{caption}
\usepackage{cite}
\usepackage{amsthm,bm}
\usepackage{amssymb}
\usepackage{circuitikz}
\usepackage{diagbox}
\usepackage{tikz}
\usepackage{subfig}    
\usepackage{blindtext, graphicx}   
\usepackage{amsmath}   
\usepackage{multicol,lipsum}
\usepackage{amssymb}   
\usepackage{lipsum} 
\usepackage{amsfonts}   
\usepackage{lipsum}   
\usepackage{breqn}
\usepackage{float}   
\usepackage{cuted}     
\usepackage{amsthm}   
\usepackage{pst-solides3d}
\usepackage{float}
\usepackage{breqn}
\usetikzlibrary{hobby}
\usepackage[normalem]{ulem}
\usepackage{pgfplots}
\usepackage{pifont}
\usepackage[colorinlistoftodos,prependcaption,textsize=small]{todonotes}
\usepackage{mathtools}
\usepackage{siunitx}
\DeclarePairedDelimiterXPP\BigOSI[2]%
  {\mathcal{O}}{(}{)}{}%
  {\SI{#1}{#2}}

\newcommand{\J}{\mathbf{J}}
\newcommand{\I}{\mathbf{I}}
\newcommand{\A}{\mathbf{A}}

\newcommand{\C}{\mathbf{C}}

\newcommand{\E}{\mathbf{E}}
\newcommand{\R}{\mathbf{R}}
\newcommand{\X}{\mathbf{X}}

\newcommand{\M}{\mathbf{M}}
\newcommand{\U}{\mathbf{U}}

\newcommand{\Z}{\mathbf{Z}}

\newcommand{\rr}{\mathbf{r}}

\newcommand{\sub}[1]{\bar{#1}}
\newcommand{\Region}{\varOmega}
\newcommand{\SubRegion}{\sub{\Region}}

\begin{document}
\title{Fundamental Limits on Substructure Dielectric Resonator Antennas}
\author{Binbin~Yang,~\IEEEmembership{Member,~IEEE}, Jaewoo Kim, and~Jacob~J.~Adams,~\IEEEmembership{Senior Member,~IEEE}

	\thanks{B. Yang is with the Department of Electrical and Computer Engineering, North Carolina Agricultural and Technical State University, Greensboro, NC, USA (email: byang1@ncat.edu).}

	\thanks{J. Kim and J. J. Adams are with the Department of Electrical and Computer Engineering, North Carolina State University, Raleigh, NC, USA (e-mail: jacob.adams@ncsu.edu).}
}

\maketitle

\begin{abstract}
We show theoretically that the characteristic modes of dielectric resonator antennas (DRAs) must be capacitive in the low frequency limit, and show that as a consequence of this constraint and the Poincar\'{e} Separation Theorem, the modes of any DRA consisting of partial elements of an encompassing super-structure cannot resonate at a frequency that is lower than that of the encompassing structure. Thus, design techniques relying on complex sub-structures to miniaturize the antenna, including topology optimization and meandered windings, cannot apply to DRAs. Due to the capacitive nature of the DRA modes, it is also shown that the Q factor of any DRA sub-structure will be bounded from below by that of the super-structure at frequencies below the first self-resonance of the super-structure. We demonstrate these bounding relations with numerical examples. 

\end{abstract}


\section{Introduction}

\IEEEPARstart{D}{ielectric} resonator antennas (DRAs) have drawn significant attention from researchers and engineers since their initial investigation by Long \cite{long1983}. One of the most appealing features of DRAs is their high radiation efficiency due to lack of conductive loss. DRAs remain highly efficient even at millimeter wave frequency range, whereas conductive patch antennas experience considerable degradation in radiation efficiency \cite{lai2008comparison}. Besides their high radiation efficiency, DRAs also feature compact size due to their high dielectric constant, and offer more design freedom owing to their various material options and 3-dimensional geometries. 

Though many DRA designs have been reported in literature covering various applications, such as broadband designs \cite{chair2004wideband,huang2007compact}, multi-port applications \cite{yan2011design,abdalrazik2017three}, beamforming arrays \cite{chow1995cylindrical,su2016linearly}, and millimeter wave applications \cite{zhang2019mimo,nor2016rectangular}, DRAs with complex structures are less investigated than their metallic counterparts. While metallic antennas often take complex geometries based on heuristic modification of canonical antennas or pixel-based optimization \cite{johnson1999genetic,cismasu2013antenna,hassan2014topology,ethier2014antenna,yang2016systematic}, the majority of reported DRA designs are based on canonical shapes, such as rectangular, cylindrical and spherical resonators \cite{Leung1993Theory,abdalrazik2017three,fang2014theory}. \textcolor{black}{However, there may also be performance benefits to more complex DRA geometries. For example, complex geometries are often employed to miniaturize metallic antennas \cite{best2009low,warnagiris1998performance,adams2011conformal}, or to broaden the bandwidth (reduce the Q factor) by combining multiple modes of different reactive energies \cite{capek2016optimal}. However, it remains unclear whether DRAs with complex geometry offer any such performance benefit, and if so, why such complex DRAs have not been as thoroughly studied as complex metallic structures.  A more fundamental analysis of DRA limitations would provide valuable guidance for future DRA designs.}

\textcolor{black}{The objective of this work is to understand whether DRAs are limited in a fundamentally different way from metallic antennas using both theory and numerical simulation. Recent observational studies have reported that shape optimization approaches applied to DRAs within a bounded volume appear to only increase the electrical size of the antenna \cite{dhar2013wideband,alroughani2020shape}, in direct contrast with miniaturization techniques used for metallic antennas.  Thus, we study the physical limits of DRAs and reveal the unique constraints on DRAs. Specifically, in Section II, we mathematically show that DRAs support only capacitive characteristic modes in the low frequency limit, a distinctive behavior compared to metallic antennas. In Section III, leveraging the capacitive nature of DRAs' characteristic modes and the Poincar\'{e} Separation Theorem, it is further established that a DRA cannot be made resonant at a lower frequency by removing selected parts from an encompassing dielectric super-structure, excluding the benefit of antenna miniaturization through complex shape design, e.g. meandering, helix or similar methods that are widely adopted for small metallic antennas. The capacitive nature of DRAs' characteristic modes also excludes the possibility of Q factor minimization by mode combining below the fundamental resonance of the superstructure, resulting into an additional bound on modal Q factors. Section IV validates the bounding relations on resonance frequencies and Q factors through numerical examples, and Section V discusses the implications of these findings on DRA miniaturization.}

\section{Characteristic Modes of DRAs}
\label{sec:capactive_nature}

The electric field integral equation (EFIE) in a dielectric object using the volume equivalence principle is:
\begin{equation}
    \frac{\J(\rr)}{j\omega\epsilon_0(\epsilon_r(\rr)-1)}+j\omega \A(\J) + \nabla \Phi(\J)=\E^i(\rr)
\end{equation}
where the polarization current $\J(\rr)$ is related to the total electrical field as $\J(\rr) = j\omega\epsilon_0(\epsilon_r(\rr)-1)\E(\rr)$; $\A(\J)$ and $\nabla \Phi(\J)$ are the magnetic vector potential and the electric scalar potential, $\E^i(\rr)$ is the incident field, and $\epsilon_r(\rr)$ is the relative permittivity distribution of the dielectric object. 

With proper basis expansion, the above equation can be converted into a matrix equation using the method of moments (MoM) \cite{SWG1984,zhang2014solving}: 
\begin{equation}
    \Z\I=(\R+j\X)\I=\mathbf{V}^i
\end{equation}
where $\X$ and $\R$ are the imaginary and real parts of the antenna's MoM impedance matrix $\Z$, $\I$ is the basis coefficient vector, and $\mathbf{V}^i$ is the excitation vector. 

The characteristic modal analysis (CMA) uses these matrices to formulate the generalized eigenvalue equation \cite{harrington1971theory,Harrington1972Dielectric}:
\begin{equation}
\label{eq:CMA_eq}
    \X \I_n=\lambda_n \R \I_n,
\end{equation}
\noindent in which the characteristic currents $\I_n$ produce far field orthogonal patterns and the characteristic eigenvalues $\lambda_n$ are related to the net stored magnetic and electric energy of the antenna \cite{harrington1971theory}. The modes with negative eigenvalues are capacitive and those with positive eigenvalues are inductive.  The eigenvalues vary with frequency, and when their magnitudes become very small (approaching resonance), they dominate the radiation response of the antenna.  Thus, for antennas that are near their fundamental resonance, a small number of modes are often sufficient to fully describe the antenna's response.  In most CMA studies, the antennas under investigation consist of conductive structures and both capacitive and inductive modes are observed below the fundamental resonance.

However, the characteristic modes of DRAs differ in a fundamental way from the metallic antennas. It has recently been observed that a DRA displays only capacitive characteristic modes \cite{Alroughani2014Observations,huang2018investigations,Miers2016Computational}, but no mathematical proof of this observation has been reported. Here in this section, we will establish a mathematical proof of the capacitive nature of the characteristic modes of DRAs, allowing for new analysis of the physical limits of DRAs in Section \ref{section:theory}.

\subsection{Capacitive Nature of the DRA Modes}
The stored electric energy ($W_e^{vac}$), the stored magnetic energy ($W_e^{vac}$) and the radiated power of an antenna in free space can be written in terms of its current and charge density ($\J(\rr)$, $\nabla \cdot \J(\rr)$) respectively as \cite{gustafsson2016antenna,vandenbosch2010reactive,capek2012method}: 
\begin{dmath}
\label{eq:We_vac}
 W_e^{vac}=  \frac{1}{16\pi \omega^2\epsilon_0} \bigg( \int_{\Omega_1}  \int_{\Omega_2} \nabla_1 \cdot \J_1(\rr_1)\nabla_2 \cdot \J_2^*(\rr_2) \frac{\cos(k r_{21})}{r_{21}} d\Omega_1 d\Omega_2 - 
 \frac{k}{2} \int_{\Omega_1}  \int_{\Omega_2} \left[ k^2\J_1(\rr_1) \cdot \J_2^*(\rr_2)-\nabla_1 \cdot \J_1(\rr_1)\nabla_2 \cdot \J_2^*(\rr_2)\right]\sin(k r_{21})d\Omega_1 d\Omega_2 \bigg)
=\frac{1}{4\omega}\I^H\X_e^{vac}\I
\end{dmath}

\begin{dmath}
\label{eq:Wm_vac}
 W_m^{vac}=  \frac{1}{16\pi \omega^2\epsilon_0} \bigg( k^2 \int_{\Omega_1}  \int_{\Omega_2} (\J_1(\rr_1) \cdot \J_2^*(\rr_2)) \frac{\cos(k r_{21})}{r_{21}} d\Omega_1 d\Omega_2 - 
 \frac{k}{2} \int_{\Omega_1}  \int_{\Omega_2} \left[ k^2\J_1(\rr_1) \cdot \J_2^*(\rr_2)-\nabla_1 \cdot \J_1(\rr_1)\nabla_2 \cdot \J_2^*(\rr_2)\right]\sin(k r_{21})d\Omega_1 d\Omega_2 \bigg)
 =\frac{1}{4\omega}\I^H\X_m^{vac}\I
\end{dmath}

\begin{dmath}
\label{eq:P_rad}
P_{rad}=  \frac{1}{8\pi \omega\epsilon_0} \int_{\Omega_1}  \int_{\Omega_2} \left[ k^2  (\J_1(\rr_1) \cdot \J_2^*(\rr_2))-(\nabla_1 \cdot \J_1(\rr_1))(\nabla_2 \cdot \J_2^*(\rr_2))\right]\frac{\sin(k r_{21})}{r_{21}}d\Omega_1 d\Omega_2 
 =\frac{1}{2}\I^H\R\I
\end{dmath}
where $\Omega$ is the DRA domain, $k=\omega/c$ is the phase constant, with $\omega$ being the angular frequency and $c$ being the speed of light in free space; $r_{21}=|\rr_2-\rr_1|$ is the distance between sources, and $\X_e^{vac}$ and $\X_m^{vac}$ are the matrix operators for the stored electric and magnetic energy in free space.

In a volume equivalent problem featuring only dielectric bodies, the antenna sources in \eqref{eq:We_vac}-\eqref{eq:P_rad} are the equivalent polarization currents and charges in free space. Moreover, another term related to the electric energy stored in the dielectric materials must be added as \cite{vandenbosch2010reactive}:
\begin{equation}
\label{eq:We_mat}
     W_e^{mat}=  \frac{1}{4\omega^2 \epsilon_0 } \int_{\Omega} \frac{\J(\rr) \cdot \J^*(\rr)}{\epsilon_r(\rr)-1}d\Omega
     =\frac{1}{4\omega}\I^H\X_e^{mat}\I
\end{equation}

The power, energy and the MoM matrix of a DRA antenna system are related as\cite{gustafsson2016antenna}:
\begin{dmath}
   \frac{1}{2} \I^H\Z\I=\frac{1}{2}\I^H\R\I+j\frac{1}{2}\I^H\X\I=\frac{1}{2}\I^H\R\I+j\frac{1}{2}\I^H(\X_m^{vac}-\X_e^{vac}-\X_e^{mat})\I=P_{rad}+j2\omega (W_m^{vac}-W_e^{vac}-W_e^{mat})
\end{dmath}

To prove that the eigenvalues of a DRA must be negative in the low frequency limit, we examine the limiting behavior of the energy terms in \eqref{eq:We_vac}-\eqref{eq:We_mat}.  First, we take the the first order limit that $\cos(kr)\approx 1$ and $\sin(kr)\approx kr$ as $kr\rightarrow 0$, and follow the general current expansion in \cite{gustafsson2012physical} that $\J=\J^{(0)}+k\J^{(1)}+o(k)$ as $k\rightarrow 0$, where $\nabla\cdot \J^{(0)}=0$ and $\nabla \cdot \J^{(1)}=-jc\rho$, with $\rho$ being the charge density.
However, because the polarization current in dielectric objects is related to the electric field as $\J = j\omega\epsilon_0(\epsilon_r-1)\E$ (in contrast with the conduction current $\J=\sigma\E$), there is no frequency independent current in lossless DRAs, namely $\J^{(0)}=0$. Now with the expansion $\J=k\J^{(1)}+o(k)$ for polarization currents, it can be shown that as $k\rightarrow 0$, the three energy terms in (\ref{eq:We_vac}),(\ref{eq:Wm_vac}) and (\ref{eq:We_mat}) are asymptotically related to $k$ as \cite{gustafsson2012physical,vandenbosch2011simple,geyi2003method}:

\begin{dmath}
\label{eq:Wm_vac_asym}
    W_m^{vac}\approx \frac{ k^4}{16\pi \omega^2\epsilon_0} \int_{\Omega_1}  \int_{\Omega_2} \left [ \frac{\J_1^{(1)}(\rr_1) \cdot \J_2^{(1)*}(\rr_2)}{r_{21}}+\frac{c^2}{2} \rho_1(\rr_1) \rho_2^*(\rr_2)r_{21}\right ] d\Omega_1 d\Omega_2 
 \approx C_1 k^2 \\
\end{dmath}

\begin{dmath}
\label{eq:We_vac_asym}
    W_e^{vac} \approx \frac{c^2 k^2}{16\pi \omega^2\epsilon_0} \int_{\Omega_1}  \int_{\Omega_2}   \frac{\rho_1(\rr_1) \rho_2^*(\rr_2)}{r_{21}} d\Omega_1 d\Omega_2  \approx C_2 k^0 \\
\end{dmath}

\begin{dmath}
\label{eq:We_mat_asym}
    W_e^{mat}\approx\frac{k^2}{4\omega^2 \epsilon_0 } \int_{\Omega} \frac{\J^{(1)}(\rr) \cdot \J^{(1)*}(\rr)}{\epsilon_r(\rr)-1}d\Omega \approx C_3 k^0
\end{dmath}

Due to the non-negative nature of these energy terms, $C_1, C_2$, and $C_3$ are non-negative constants related to the antenna geometry and source distribution\footnote{Though these fomulations can produce negative energy for certain electrically large antennas \cite{gustafsson2012physical,schab2018energy}, the problem is not present for electrically small antennas, which is the case when $k\rightarrow 0$.}.
Recalling the relation $\I^H\X\I=4\omega (W_m^{vac}-W_e^{vac}-W_e^{mat})$, it follows that 
\begin{equation}
\label{eq:negative_definite}
    \I^H\X\I<0 \;\; \mathrm{as}\;\; k\rightarrow 0.
\end{equation}

\textbf{Because the above relation is valid for all possible current distributions on the antenna, \eqref{eq:negative_definite} implies that the $\X$ matrix is negative definite as $k\rightarrow 0$. Recalling that the $\R$ matrix is positive definite\footnote{Here we are treating $\R$ matrix as positive definite.The case when $\R$ matrix is positive semidefinite can be treated equally well by introducing a small amount of loss as in \cite{schab2018lower}.} in general for antennas ($\I^H\R\I=P_{rad}> 0$), and the fact that the characteristic eigenvalue is equivalent to the Rayleigh quotient as \cite{capek2016analytical,chen2015characteristic}: 
\begin{equation}
    \lambda_n=\frac{\I^H_n\X\I_n}{\I_n^H\R\I_n}
    \label{eq:Rayleigh_quotient}
\end{equation}
the relation in (\ref{eq:negative_definite}) means that all the characteristic eigenvalues of DRAs are negative at low frequencies, making them exclusively capacitive modes.}

When metallic structures supporting frequency independent, loop-type currents ($\J^{(0)}\neq 0$) are involved, the magnetic energy term ($W_m^{vac}$) due to $\J^{(0)}$ will be proportional to $k^0$, no longer guaranteeing the negative definiteness of the $\X$ matrix and allowing the possibility of inductive modes.

\section{Physical Limits on the Resonant Frequencies and Q Factor of Dielectric Resonator Antennas}
\label{section:theory}
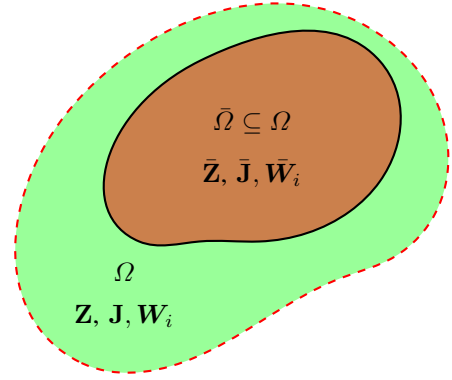
\begin{figure}
\begin{center}
\begin{tikzpicture}[use Hobby shortcut,closed=true]	
	
	\draw[red,thick,dashed, fill=green, fill opacity = 0.4,scale=1.4] (0,0) .. (0,2.5).. (3,3).. (3,1).. (2.5,0.8).. (1,0).. (0,0);

	\draw[closed = true, thick, fill=red, fill opacity = 0.5, scale=0.8] (0+1,0+2) .. (0+2,3+2).. (3+2,3+2).. (3+2,1+2).. (2.5+2,0.5+2).. (1+2,0+2).. (0+2,0+2);
	\node at (2.3,3.2){$\SubRegion\subseteq\Region$};
	\node at (2.3,2.5){$\bar{\mathbf{Z}}$, $\bar{\mathbf{J}},\sub{\boldsymbol{W}_i}$};
	\node at (0.6,1.2){$\Region$};
	\node at (0.6,0.6){$\mathbf{Z}$, $\mathbf{J},\boldsymbol{W}_i$};
\end{tikzpicture}
\caption{A 3D region $\Region$ and an arbitrary 3D subregion $\SubRegion \subseteq \Region$. By \eqref{eq:WSubset}, functions with support confined to the subregion $\SubRegion$ can be expressed in either basis ($\mathbf{W}_i$ or $\mathbf{\bar{W}}_i$).}
\label{fig:diagram}
\end{center}
\end{figure}

\textcolor{black}{The capacitive nature of DRAs' characteristic modes have some unique implication on the physical parameters (resonance frequency, Q factor, etc.) of DRAs as revealed through the analysis of DRA substructures. Substructure antennas are antenna structures that are obtained by combining parts of an encompassing super-structure as illustrated in Fig.~\ref{fig:diagram}.} This is often the case when designing antennas in a confined region, \textit{e.g.,} user equipment with limited space. The bounds of metallic substructure antennas have been studied in \cite{schab2018lower}, where it is mathematically shown that the Q factors of a substructure metallic antenna are bounded by those of its encompassing super-structure. 
Here, with the conclusions from the previous section, we investigate the bounds of dielectric substructure antennas.

The definition of substructure antennas for DRAs is similar to their metallic counterparts in \cite{schab2018lower}. Consider a 3D dielectric super-structure $\mathbf{\Omega}$ with the basis set $\{\mathbf{W}_i\}$ on which any arbitrary dielectric polarization current can be expanded as $\J=\sum I_i \mathbf{W}_i$, as illustrated in Figure \ref{fig:diagram}.
Similarly, we consider a substructure $\mathbf{\bar{\Omega}}\subseteq \mathbf{\Omega}$ with its basis set as $\{\mathbf{\bar{W}}_i\}$, and any arbitrary current on it can be expanded as $\bar{\J}=\sum \bar{I_i} \bar{\mathbf{W}_i}$. It follows that
\begin{equation}
\label{eq:WSubset}
    \mathrm{span}\{\mathbf{\bar{W}}_i\}\subseteq \mathrm{span}\{\mathbf{W}_i\}. 
\end{equation}
We here consider a finite dimension problem, where dim$\{\mathbf{W}_i\}=K$, and dim$\{\mathbf{\bar{W}}_i\}=\bar{K}$.

\textcolor{black}{Useful mathematical relations can be derived between the impedance matrices, energy matrices, and eigenvalues of the super-structure and substructure DRAs \cite{schab2018lower}. For completeness, these relationships are given in detail in Appendix A.}

\subsection{Bound on Characteristic Eigenvalues and Resonant Frequencies of Substructure DRAs}
\label{sec:eigval_bound}

\textcolor{black}{As shown in Appendix A, the characteristic eigenvalues of the super-structure ($\lambda$) and its arbitrary substructure ($\bar{\lambda}$) are related as:}

\begin{equation}
\label{eq:PSTResult_lambda}
\lambda_k  \geqslant \bar{\lambda}_k \geqslant \lambda_{k+K-\bar{K}},  1\geqslant k\geqslant \bar{K}.
\end{equation} 
for all frequencies, where $K$ is the rank of the super-structure Z matrix, and $\sub{K}$ is the rank of the substructure matrices. 
\textbf{The key interpretation of the bounding relation in (\ref{eq:PSTResult_lambda}) is that each $k^{\mathrm{th}}$ characteristic eigenvalue of any substructure DRA is bounded from above by the $k^{\mathrm{th}}$ characteristic eigenvalues of the super-structure.} Note that here the modes are ordered purely based on the algebraic value of the eigenvalues. This bounding relation is valid for both metallic and dielectric antennas, but due to the capacitive nature of DRA modes, this bounding relation provides new insight on the performance limits of DRAs, as elaborated below.

The resonant frequency of each characteristic mode is defined as the frequency at which its eigenvalue reaches zero. Since we have shown in Section \ref{sec:capactive_nature} that small DRAs have only capacitive modes, the implication of the bounding relation in (\ref{eq:PSTResult_lambda}) is that all the substructure DRAs will have eigenvalues that are lower (more negative) than the super-structure. Specifically, if we look at the eigenvalue with the lowest resonant frequency, the bound in (\ref{eq:PSTResult_lambda}) implies that the first eigenvalue of all substructure DRAs that fit within the super-structure must cross zero at a frequency that is equal to or higher than that of the super-structure. This is true for not just the first, but the higher order characteristic modes as well. Namely,
\begin{equation}
\label{eq:PSTResult_freq}
f_k  \leqslant \bar{f}_k \leqslant f_{k+K-\bar{K}}, 1\geqslant k\geqslant \bar{K}.
\end{equation} 
where $f_k$ and $\bar{f}_k$ are the resonant frequencies of the $k$-th characteristic mode for the complete and the substructure DRAs respectively.
\textbf{In other words, if arranging the characteristic modes based on the algebraic value of the eigenvalues, the characteristic modal resonance frequencies of all the substructure antennas are bounded below by those of the super-structure. } Consequently, DRAs cannot be made resonant at a frequency lower than the original resonant frequency of the super-structure by optimizing the substructure antenna geometry. This result provides valuable guidance for future DRA design, optimization and synthesis.

\subsection{Bound on Q Factor of Substructure DRAs}
\label{sec:Q_bound}
In metallic antennas supporting both inductive and capacitive modes at low frequencies, it is possible to combine the inductive and capacitive modes to further lower the antenna Q factor \cite{capek2016optimal,thal2006new}. However, DRA modes cannot be inductive below their fundamental resonance and are therefore destined to store more electric energy than magnetic energy below the first modal resonance ($f_1$).  Thus, the tuned Q factor below the first modal resonance for DRAs can be written as \cite{collin1964evaluation}:
\begin{dmath}
    Q=\frac{2\omega\; \mathrm{max}\{W_m^{vac},W_e^{vac}+W_e^{mat}\}}{P_{rad}}=\frac{2\omega\; (W_e^{vac}+W_e^{mat})}{P_{rad}}=\frac{\I^H(\X_e^{vac}+\X_e^{mat})\I}{\I^H \R\I}
\end{dmath}
The minimization of the above Rayleigh quotient can be reformulated as the following eigenvalue equation,
\begin{equation}
\label{eq:Q_eigen_eqn}
    (\X_e^{vac}+\X_e^{mat}) \I_n=Q_n\R \I_n
\end{equation}
where $\X_e^{vac}$ and $\X_e^{mat}$ are the energy operators defined in (\ref{eq:We_vac}) and (\ref{eq:We_mat}).

We can rewrite this equation in a form similar to (\ref{eq:SubRegionGEP2}) by invoking the basis transformation between the substructure and super-structure DRAs. Arranging the eigenvalues (in this case, $Q_n$) in ascending order, and invoking Poincar\'{e} Separation Theorem \cite{Horn2012} (see Appendix B), we can again show the following bounding relation:
\begin{equation}
\label{eq:PSTResult_Q}
Q_k  \leqslant \bar{Q}_k \leqslant Q_{k+K-\bar{K}}, 1\geqslant k\geqslant \bar{K},
\end{equation} 
for $f\leq f_1$. Namely, before the first modal resonant frequency of the super-structure, the Q factor of all substructure DRAs are bounded below by that of the complete DRA structure \textcolor{black}{not only for the fundamental mode but also for the higher order modes.}

\section{Numerical Examples}
\label{section:Numerical_examples}
In this section, we illustrate the theoretical bound established in Section \ref{section:theory} through several numerical examples.

\subsection{Optimization of DRA Eigenvalues}
\label{section:eigenvalue_optimization}
We select a rectangular DRA as our example problem to demonstrate the eigenvalue and Q factor bounds. The DRA has a dimension of $40\times25\times16$ mm$^3$ and is placed in free space. Considering the variety of materials in DRA designs, two different materials are investigated here: alumina with permittivity of 9.8 and zirconia with permittivity of 23 \cite{oh2019microwave}. Loss is ignored for the characteristic modal analysis (CMA). An in-house volume EFIE method of moments code \cite{yang2017quality} based on the SWG basis \cite{SWG1984} and characteristic mode solver in \cite{adams2013broadband} are used for the CMA.  This code has been previously validated against both analytical calculations and measurements, as well as commercial CMA codes, such as FEKO.

\begin{figure}[t]
\centering%
\subfloat[]{%
\centering
\includegraphics[width=0.45\linewidth]{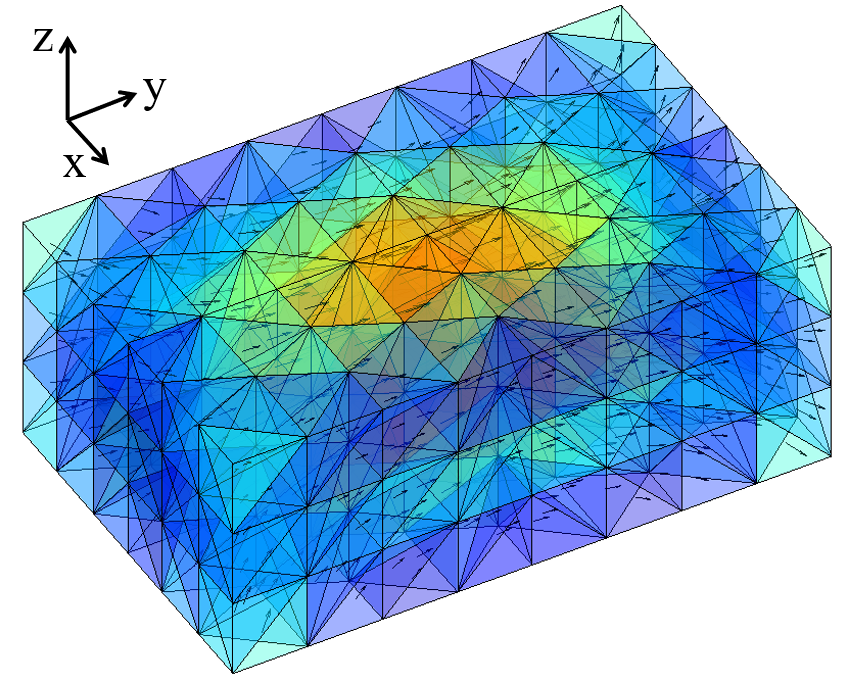}
\label{fig:current_mode1}
}%
\subfloat[]{%
\centering
\includegraphics[width=0.45\linewidth]{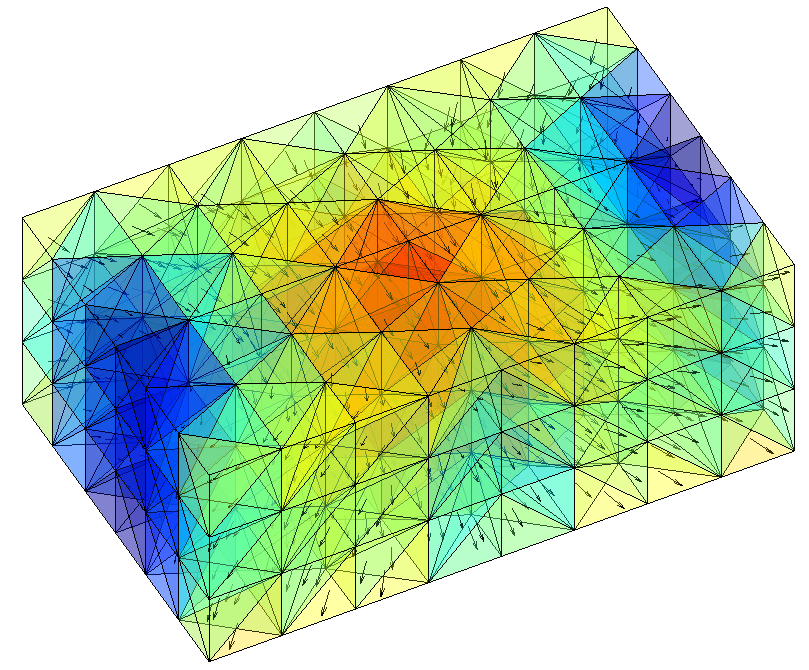}
\label{fig:current_mode2}
}%
\\
\subfloat[]{%
\centering
\includegraphics[width=0.45\linewidth]{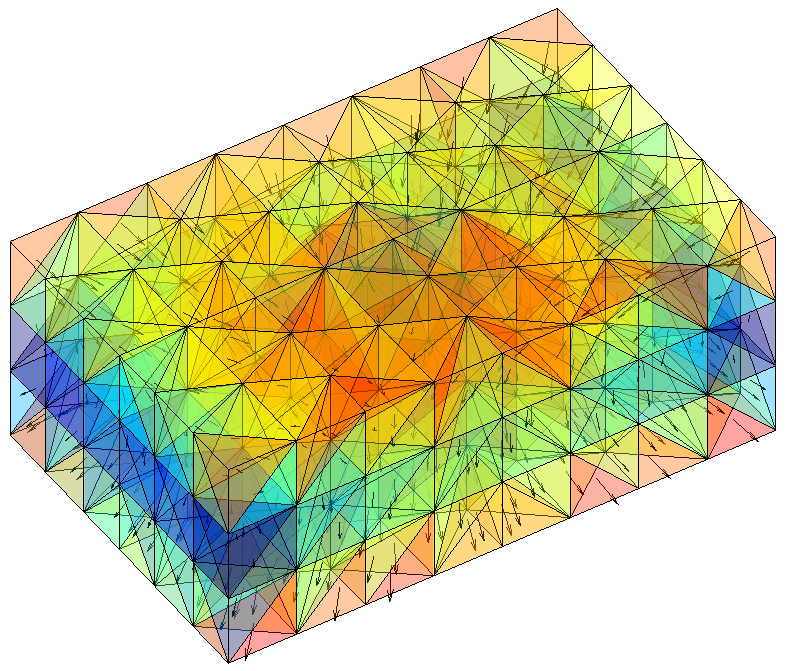}
\label{fig:current_mode3}
}%
\subfloat[]{%
\centering
\includegraphics[width=0.45\linewidth]{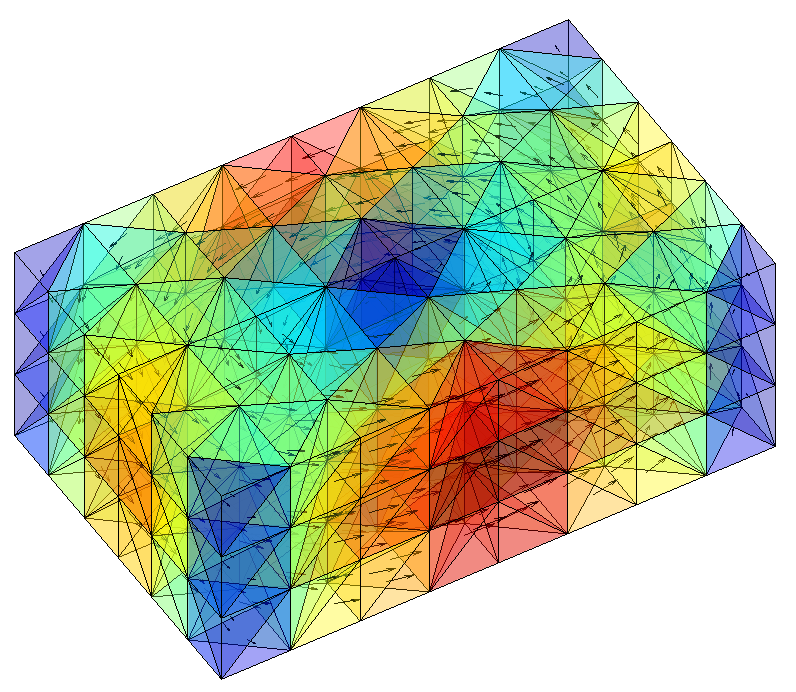}
\label{fig:current_mode4}
}%
\\
\caption{\textcolor{black}{The characteristic modal currents (polarization currents) of the studied rectangular DRA with $\epsilon_r=9.8$ at 2.5 GHz: (a) mode 1 ($TM^y_{111}$); (b) mode 2 ($TM^x_{111}$); (c) mode 3 ($TM^z_{111}$) and (d) mode 4 ($TE^z_{111}$).}}
\label{fig:modal current}
\end{figure}

\textcolor{black}{The polarization current distributions of the first four characteristic modes of the rectangular DRA are given in Figure~\ref{fig:modal current}. The first three modes are roughly equivalent to electric dipoles in the y, x and z directions, and the fourth mode is similar to a magnetic dipole in the z direction. We attach modal nomenclature following \cite{mongia1992theoretical} in Figure~\ref{fig:modal current}.}

The first resonance of the super-structure is 2.94 GHz for the alumina DRA and 1.96 GHz for the zirconia DRA. The theoretical bounds in (\ref{eq:PSTResult_lambda}) and (\ref{eq:PSTResult_freq}) predict that all substructure DRAs should have smaller eigenvalues than the super-structure, and resonate at a frequency no lower than the first resonance of the super-structure. To test the derived bounds, we conduct an optimization that attempts to maximize the eigenvalues to determine whether they can exceed those of the super-structure.  

An optimization is conducted at 2.5 GHz for the alumina DRA, and at 1.2 GHz for the zirconia DRA. These frequencies are below the fundamental resonances, and the characteristic eigenvalues are all negative. \textcolor{black}{The optimization uses a binary genetic algorithm (GA) in which each binary gene in this case represents the presence or absence of a dielectric tetrahedron. In the GA procedure, the optimization initializes with a set of randomly generated structures (50\% chance of each tetrahedra being present or absent) that then evolves through tournament, crossover, and mutation steps. The structures' fitness is evaluated with a cost function and the candidates with the lowest costs in each generation are combined and mutated to generate the next generation of structures.  These new structures are fed back into the search cycle until the optimization procedure converges. In this study, 80 generations are used and each generation consists of 40 population. The mutation rate in this case is 12\%. Interested readers can refer to \cite{johnson1999genetic, yang2016systematic,yang2019shape,Binbin2017thesis} for further details on the GA optimization used here. Note that two geometric symmetry planes are assumed in the DRA calculation.}

The cost function defined for this optimization is,
\begin{equation}
    \mathrm{cost}=-\bar{\lambda} _k.
\end{equation}
where $\bar{\lambda}_k$ is the $k$-th characteristic eigenvalue of the DRA structure under evolutionary search. Due to the capacitive nature of the DRAs, this cost function is equivalent to maximizing the individual characteristic eigenvalues of the substructure DRAs, and thus, moving them nearer to resonance at the optimization frequency.

\begin{figure}
\centering%
\subfloat[]{%
\centering
\includegraphics[width=0.80\linewidth]{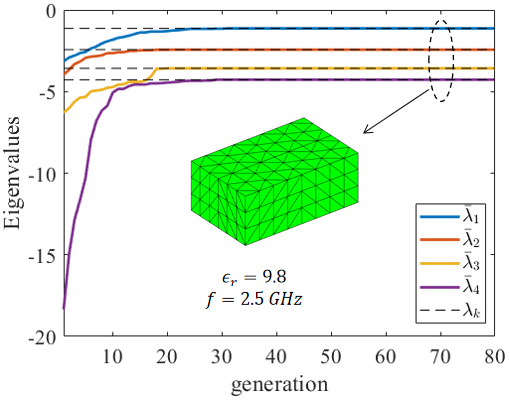}
\label{fig:lambda_convergence1}
}%
\\
\subfloat[]{%
\centering
\includegraphics[width=0.80\linewidth]{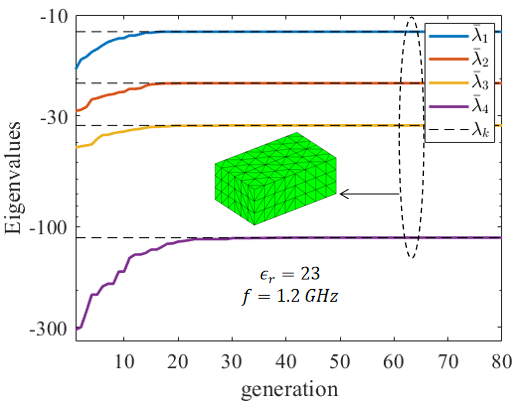}
\label{fig:lambda_convergence2}
}%
\\
\caption{(a) The first four characteristic eigenvalues of the optimized substructures (solid lines) are bounded above by those of the super-structure (dashed lines) for the alumina DRA ($\epsilon_r=9.8$) at $f=2.5$ GHz, (b) the first four characteristic eigenvalues of the optimized substructures are bounded above by that of the super-structure for the zirconia DRA ($\epsilon_r=23$) at $f=1.2$  GHz. Each individual mode is optimized independently and only the most-fit result is shown for each generation.}
\label{fig:convergence}
\end{figure}

Optimization is conducted independently for the first four characteristic modes of the studied rectangular DRAs in an attempt to move any of them above the super-structure's baseline value. Figure \ref{fig:convergence}(a) shows the convergence plot of the first four eigenvalues of the alumina DRA and Figure \ref{fig:convergence}(b) shows that of the zirconia DRA. As predicted by the theoretical result in Equation (\ref{eq:PSTResult_lambda}), all the four eigenvalues of the substructures are bounded above by those of the super-structure. While each optimization is initialized randomly, and thus starts with a relatively low initial eigenvalue, the optimization quickly pushes the best eigenvalues to the upper bounds. The final optimized geometry in all cases converges to the super-structure, confirming the bounding relation in (\ref{eq:PSTResult_lambda}). As pointed out in (\ref{eq:PSTResult_freq}), the eigenvalue bound implies a bound on the characteristic resonance frequencies.

\subsection{Numerical Optimization - Q Factor Bound}
To validate the Q factor bound established in (\ref{eq:PSTResult_Q}), we study the same DRAs studied in Section \ref{section:eigenvalue_optimization}. In this case, the cost function is defined as the modal Q factor, namely:
\begin{equation}
    \mathrm{cost}=\bar{Q}_k,
\end{equation}
where $\bar{Q}_k$ is the $k$-th lowest modal Q factor of the DRA structure under evolutionary search. \textcolor{black}{The calculation of the modal Q factors of the DRA is conducted based on the energy operators discussed in Section~\ref{sec:capactive_nature} and the validation of the method against results in literature can be found in \cite{yang2017quality}.} To be consistent, the optimization frequencies remain 2.5 GHz for the alumina DRA and 1.2 GHz for the zirconia DRA. Based on the result in (\ref{eq:PSTResult_Q}), the lowest $k$-th modal Q factor of the complete DRA structure should bound the lowest $k$-th modal Q factor of all its substructure DRAs. Thus, additional geometric complexity cannot lower the Q factor of a DRA mode.

\begin{figure}
\centering%
\subfloat[]{%
\centering
\includegraphics[width=0.80\linewidth]{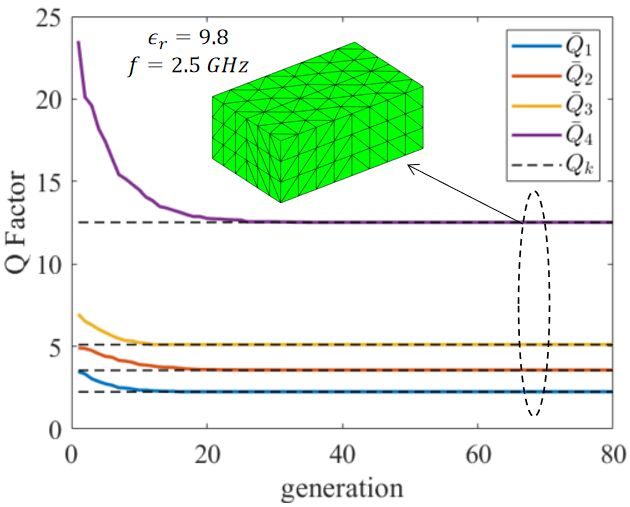}
\label{fig:Q_convergence1}
}%
\\
\subfloat[]{%
\centering
\includegraphics[width=0.80\linewidth]{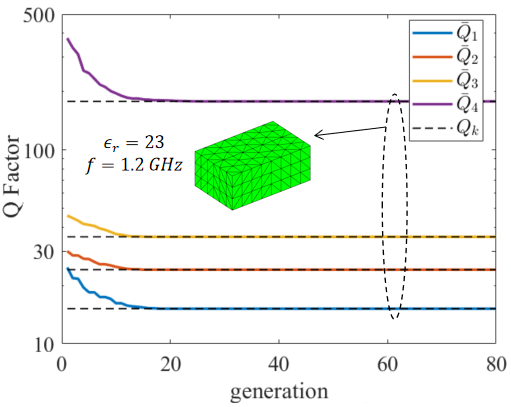}
\label{fig:Q_convergence2}
}%
\\
\caption{(a) The first four modal Q factors of the optimized substructures (solid lines) are bounded below by that of the super-structure (dashed lines) for the alumina DRA ($\epsilon_r=9.8$) at $f=2.5$ GHz, (b) the first four modal Q factors of the optimized substructures are bounded below by that of the super-structure for the zirconia DRA ($\epsilon_r=23$) at $f=1.2$  GHz. Each individual mode is optimized independently and only the most-fit result is shown for each generation.}
\label{fig:Q_convergence}
\end{figure}

The minimization of the first four modal Q factors are conducted independently using the same genetic algorithm. The optimized result of the alumina DRA is shown Figure \ref{fig:Q_convergence} (a) and Figure \ref{fig:Q_convergence} (b) shows that of the zirconia DRA. In both cases, all the four modal Q factors of the substructure antenna are bounded below by that of the super-structure, confirming the bounding relation in Equation (\ref{eq:PSTResult_Q}).  

Note that here we used the characteristic modal currents for the Q factor evaluation. Even though generally the eigenvectors in (\ref{eq:Q_eigen_eqn}) are not necessarily the same as the characteristic eigenvectors, the CM eigenvector set is a close orthogonal basis set with negligible cross energy terms, and has been employed for similar Q factor minimization problem \cite{capek2016optimal}.

\subsection{Particular Examples}
\begin{figure}
\centering%
\subfloat[]{%
\centering
\includegraphics[width=42mm]{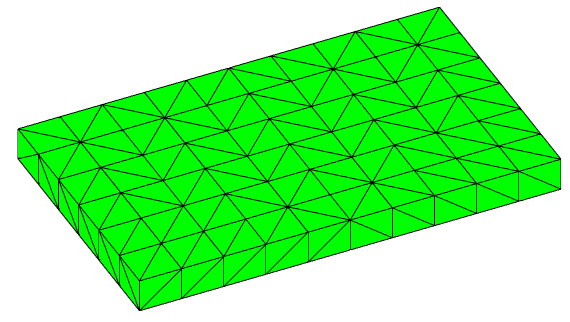}
}%
\subfloat[]{%
\centering
\includegraphics[width=42mm]{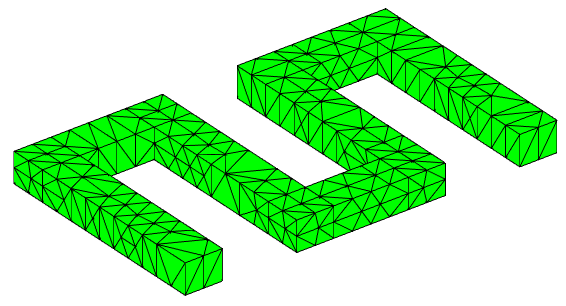}
}%
\\[0.5mm]
\subfloat[]{%
\centering
\includegraphics[width=70mm]{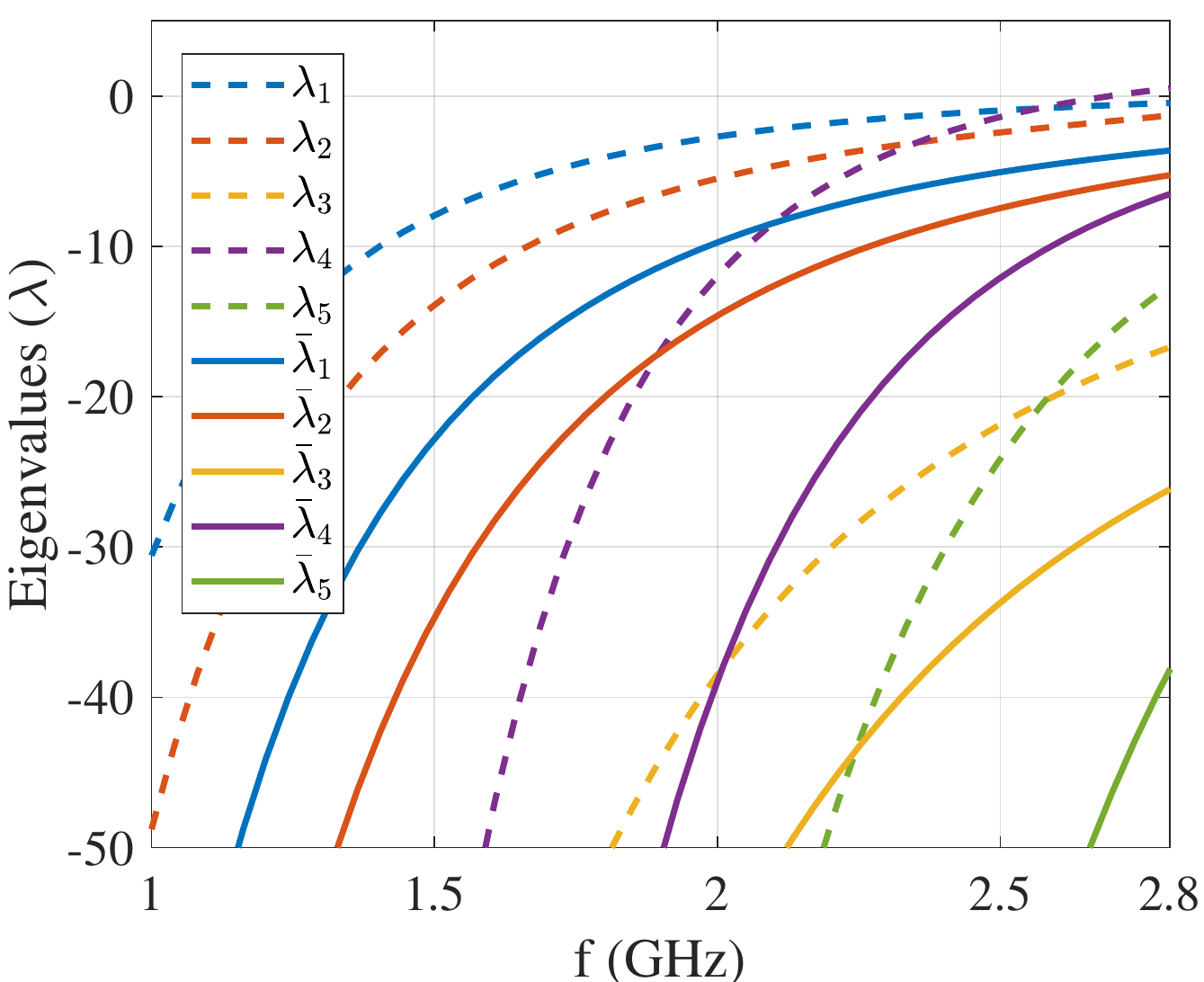}
}%
\\
\subfloat[]{%
\centering
\includegraphics[width=70mm]{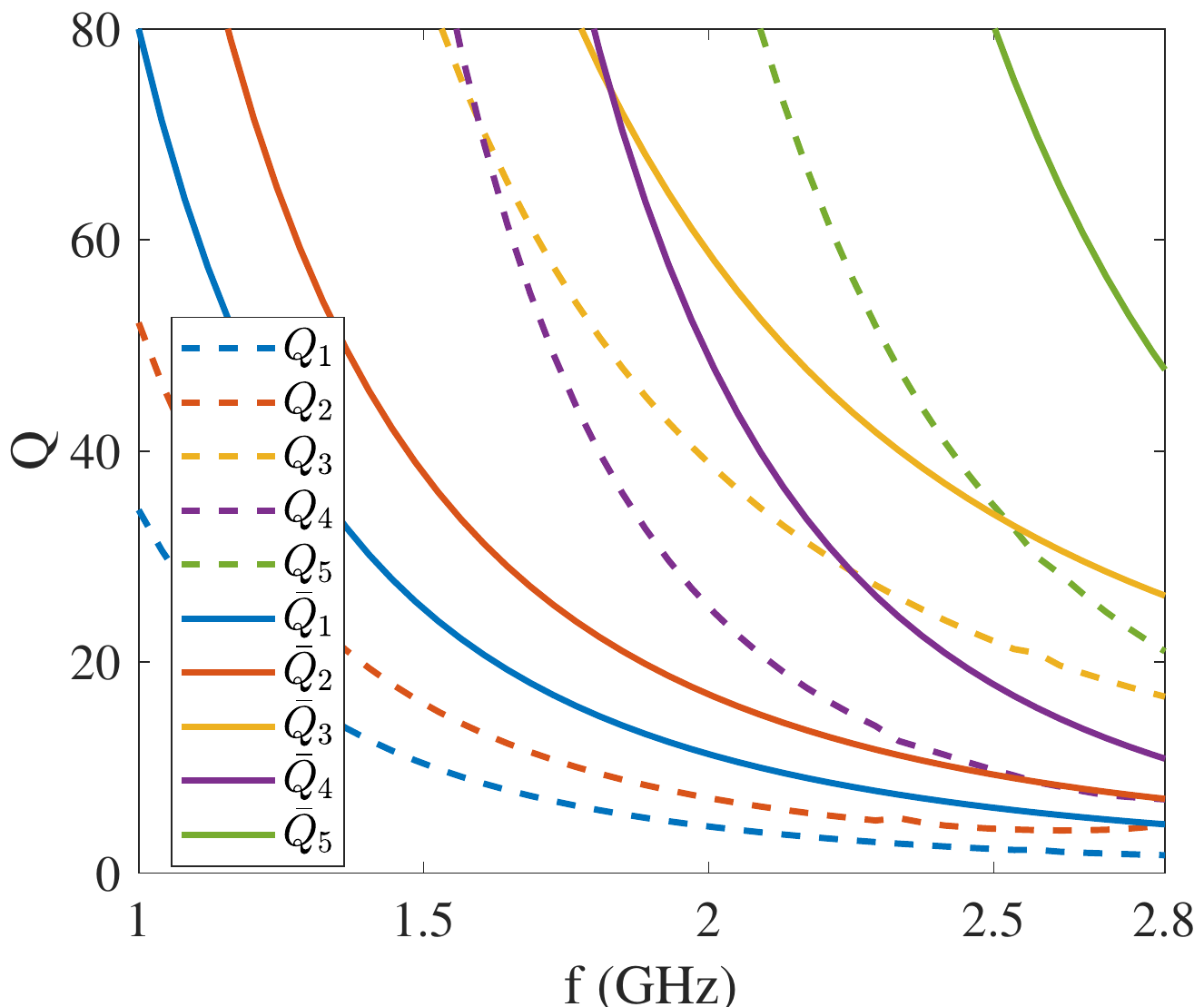}
}%
\\
\caption{(a) Sample A - super-structure, a rectangular DRA (b) Sample A - substructure, a meandered DRA that is bounded by the super-structure, (c) Comparison of the characteristic eigenvalues of the substructure (solid lines) and that of the super-structure (dashed lines), \textcolor{black}{(d) Comparison of the characteristic modal Q factors of the substructure (solid lines) and that of the super-structure (dashed lines).}}
\label{fig:Sample_A}
\end{figure}

\begin{figure}
\centering%
\subfloat[]{%
\centering
\includegraphics[width=30mm]{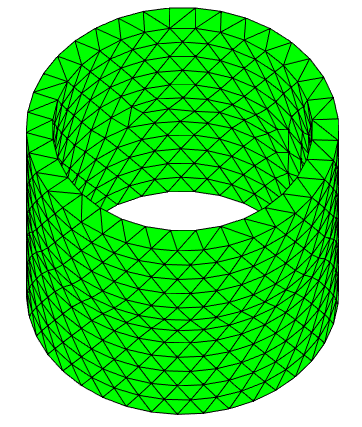}
}%
\subfloat[]{%
\centering
\includegraphics[width=30mm]{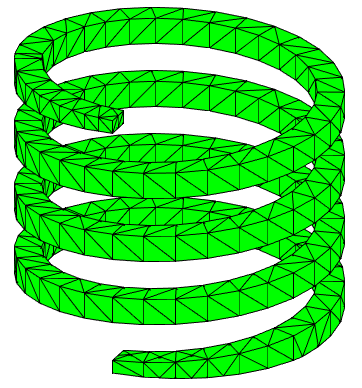}
}%
\\[0.5mm]
\subfloat[]{%
\centering
\includegraphics[width=70mm]{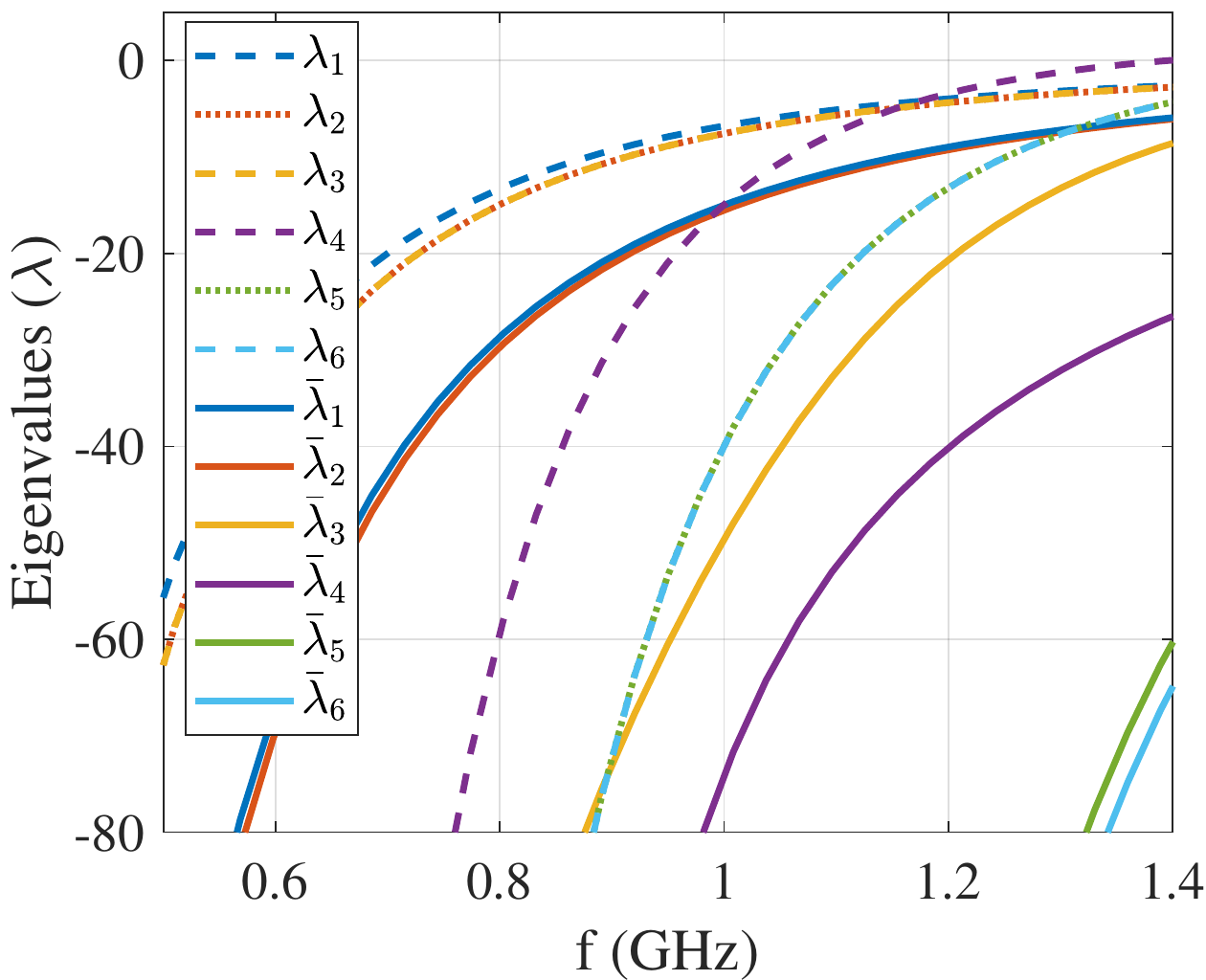}
}%
\\
\subfloat[]{%
\centering
\includegraphics[width=70mm]{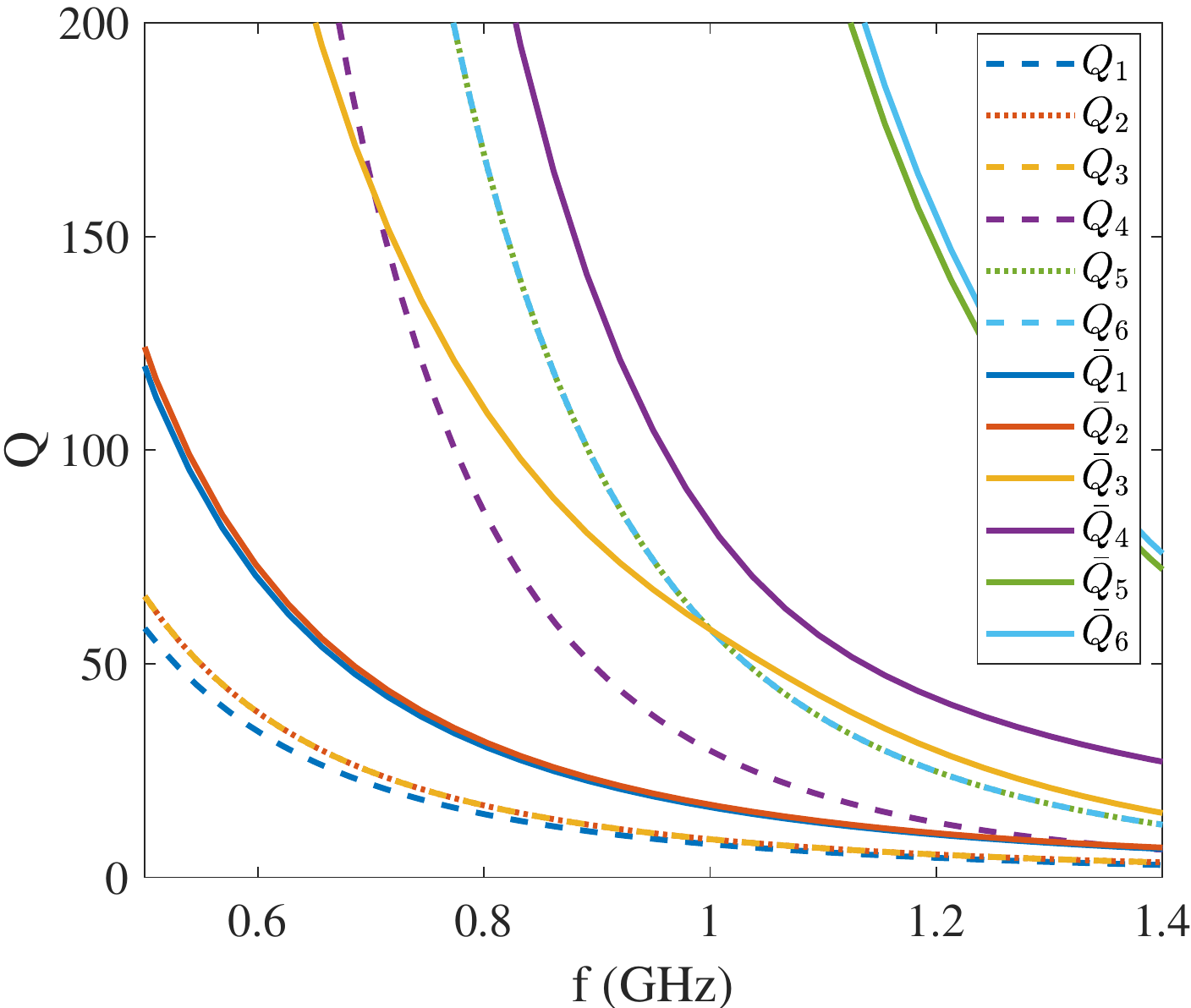}
}%
\\
\caption{(a) Sample B - super-structure, a ring DRA (b) Sample B - substructure, a helix DRA that is bounded by the super-structure, (c) Comparison of the characteristic eigenvalues of the substructure (solid lines) and that of the super-structure (dashed lines), \textcolor{black}{(d) Comparison of the characteristic modal Q factors of the substructure (solid lines) and that of the super-structure (dashed lines).}}
\label{fig:Sample_B}
\end{figure}

To better illustrate the bounding phenomena and why conventional small metallic antenna design techniques fail to work in DRA design, we provide two specific design examples. In Sample A, we use a meandering technique common in small antenna design in an attempt to lower the resonant frequency of a DRA substructure.  The meandered DRA is created within a thin, rectangular DRA block (the super-structure), as shown in Figure \ref{fig:Sample_A} (a) and (b). The rectangular DRA has a dimension of $50\times30\times4$ mm$^3$ and the meandered DRA fits within the super-structure. The strip width is 5 mm and the gap between meandered lines is 10 mm. In Sample B, a helical DRA is created within a cylindrical ring DRA (the super-structure), as shown in Figure \ref{fig:Sample_B} (a) and (b). The ring DRA has an inner radius of 25 mm and an outer radius of 30 mm. The height is 50 mm. The vertical thickness of the winding strip is 5 mm, but trimmed at the two ends to fit in the bounding ring structure. The step size of the helix winding in axis direction is 12.5 mm. Both DRAs have the permittivity of 23 (zirconia).

We calculate the eigenvalues and the resonant frequencies of both samples and their super-structures and then compare these values against the bounding relation in (\ref{eq:PSTResult_lambda}) and (\ref{eq:PSTResult_freq}). Figure \ref{fig:Sample_A}(c) compares the eigenvalue spectrum of Sample A with that of the full rectangular block, and Figure \ref{fig:Sample_B}(c) compares the eigenvalue spectrum of Sample B with that of the full ring DRA. In both Figure \ref{fig:Sample_A}(c) and Figure \ref{fig:Sample_B}(c), the eigenvalues of the super-structures are shown as dashed lines, while the eigenvalues of the substructures are shown as solid lines.  Note that in Sample B - super-structure, modes 2 and 3 are degenerate and modes 5 and 6 are degenerate. From both examples, we can see that the substructure eigenvalues are bounded above by that of the super-structure in the way defined in (\ref{eq:PSTResult_lambda}). Note that the mode numbering in the figures is not necessarily consistent across all the frequencies from the algebraic ordering perspective. 

As suggested in (\ref{eq:PSTResult_freq}), the eigenvalue bounding relation means that the resonant frequencies of both substructure antennas must be higher than those of the super-structure. Table \ref{tab:tab1} compares the resonant frequencies of both samples against their super-structures for the first four characteristic modes. It is clearly seen that the resonant frequencies of these ``electrically longer" structures in both Sample A and B actually resonate at a frequency that is higher than that of the super-structures.

\textcolor{black}{As a demonstration of the Q factor bound in \eqref{eq:PSTResult_Q}, the modal Q factors of the super- and sub-structures of Samples A and B are calculated and compared respectively in Figure \ref{fig:Sample_A} (d) and Figure \ref{fig:Sample_B} (d). Below the fundamental resonance frequencies of the super-structures (2.7 GHz for Sample A and 1.39 GHz for Sample B), all the modal Q factors of the substructures (meander DRA, helix DRA) are bounded below by that of the super-structures (rectangular DRA, ring DRA). Again, we used characteristic modal currents for the Q factor calculation as is done in Section IV B.}

\begin{table}
{
\renewcommand{\arraystretch}{1.5}
\caption{Comparison of the Characteristic Modal Resonant Frequencies in Samples A, B with Those of Their Super-structures}
\begin{center}
\begin{tabular}{|c|c|c|c|c|}
\hline
\textbf{\backslashbox{DRAs}{$\mathbf{f^{res}_m}$ (GHz)}} &  \textbf{{$f_1^{res}$}}& \textbf{$f_2^{res}$} &\textbf{$f_3^{res}$} & \textbf{$f_4^{res}$}\\
\hline
Sample A & 4.23 & 4.35 & 4.88 & 6.80\\
\hline
Super-structure A & 2.70 & 3.06 & 3.22 & 3.67\\
\hline
\hline
Sample B & 2.13 & 2.74 & 3.42 & 3.46\\
\hline
Super-structure B & 1.39 & 1.71 & 1.71 & 2.38 \\
\hline

\end{tabular}
\label{tab:tab1}
\end{center}
}
\end{table}

\section{Discussion}
\label{section:discussion}
\subsection{DRA Miniaturization}
Researchers familiar with small conductive antenna design may initially expect the meandering and helical structures in the previous section to increase the electrical length of the structure and reduce the resonant frequencies.  However, this is clearly not the case as shown in Table~\ref{tab:tab1}. The theoretical results in Section \ref{section:theory} and the numerical results presented here both agree that a DRA substructure cannot resonate at a lower frequency than its super-structure, leaving antenna designers less freedom to tune the resonance when trying to design compact DRAs. Thus, different considerations must be made for DRA miniaturization than for metallic antennas.  For example, it is common practice in metallic antenna design to optimize for a lower resonant frequency than available from the super-structure \cite{ethier2014antenna,yang2019shape}. Yet, based on the results of this work, the lack of modes with an inductive nature within the super-structure prevent the reduction of the resonant frequency of a DRA by shape optimization.  Thus, this fundamental result provides valuable guiding principles for DRA optimization and synthesis \cite{alroughani2020shape,yang2020shape}. Instead, to reduce the resonance frequencies of a DRA, one must introduce inductive energy into the antenna. This technique is often accomplished with resonant or inductive metallic structures and has been studied by several researchers. For example, in \cite{liu2016miniaturized} artificial materials based on split ring resonators are used to miniaturize a DRA. In \cite{wu2019characteristic,lee2002small}, lower resonant frequencies are created by including metallic feeding and loading structures. 

\textcolor{black}{\subsection{Multi-Mode DRAs}
Multi-mode DRAs can be used for MIMO and diversity applications. The finding in \eqref{eq:PSTResult_freq} has some important implication for MIMO DRA design and synthesis. For example, relation \eqref{eq:PSTResult_freq} shows that a DRA's modal resonant frequencies can only increase from shape manipulation confined within the superstructure. Therefore, to design or synthesize an $N$-port self-resonant MIMO DRA within a given superstructure, the desired operating frequency must be above the $N$-th modal resonance frequency of the superstructure, assuming modal resonance frequencies are ranked in ascending order. Once again, we observe that such constraints do not exist for metallic antennas as demonstrated in recent small MIMO antenna synthesis work \cite{yang2019shape}}.

\section{Conclusion}
\label{section:conclusion}
In this paper, we prove that all characteristic modes of DRAs must be capacitive in the low frequency limit. As a consequence of this constraint and the eigenvalue interlacing principle, we show that DRA modes cannot be made to resonate at a lower frequency than the resonant frequency of that mode in the encompassing super-structure.  This implies that optimization of the DRA geometry within a super-structure cannot make the antenna electrically smaller at resonance, which provides useful insight for DRA design and synthesis efforts. Moreover, we show that because no inductive modes are available for modal combining to lower the antenna Q factor, the modal Q factors of all substructure DRAs are bounded by the Q factors of the DRA super-structure at frequencies below the first modal resonance frequency of the super-structure.

\appendices
\section{Eigenvalue Bound of Substructure Antennas}

This section provides the mathematical relation between a super-structure and its sub-structures, and the bounding relation of their characteristic eigenvalues. As shown in \cite{schab2018lower,kundu2016information} using basis transformation, the super-structure and its substructure has following relations:
\begin{equation}
    \bar{\Z}=\M^H\Z\M
\end{equation}

Therefore, 
\begin{equation}
\label{eq:X_relation}
    \bar{\X}=\M^H\X\M
\end{equation}
\begin{equation}
    \bar{\R}=\M^H\R\M
\end{equation}
where $\M$ is the basis transformation matrix that relates the current coefficient as $\I=\M\sub{\I}$, \textcolor{black}{and $\Z$ and $\bar{\Z}$ are the MoM Z matrix of the super-structure and the substructure as illustrated in Figure \ref{fig:diagram}.} The transformation relation is valid for energy operators, $\X_e^{vac}$, $\X_m^{vac}$, and $\X_e^{mat}$ as well\cite{schab2018lower}.

The characteristic modal eigenvalue equation for the super-structure can be written as (\ref{eq:CMA_eq}). For a substructure antenna, the corresponding CMA equation will be 
\begin{equation}
    \sub{\X} \sub{\I}_n=\sub{\lambda}_n \sub{\R} \sub{\I}_n,
\end{equation}

Following similar steps to \cite{schab2018lower}, we can rewrite the above generalized eigenvalue equation as a new eigenvalue equation below: 
\begin{equation}
\label{eq:SubRegionGEP2}
\sub{\R}^{-\frac{1}{2}} \sub{\X} \sub{\R}^{-\frac{1}{2}} (\sub{\R}^{\frac{1}{2}} \sub{\I})=\sub{\lambda}(\sub{\R}^{\frac{1}{2}} \sub{\I})
\end{equation}

Defining the new matrix for the substructure eigenvalue problem as 
\begin{equation}
\label{eq:Cbar1}
\sub{\C}=\sub{\R}^{-\frac{1}{2}} \sub{\X} \sub{\R}^{-\frac{1}{2}}
\end{equation}
and the counterpart for the super-structure as $\C=\R^{-\frac{1}{2}} \X \R^{-\frac{1}{2}}$, it can be shown that \cite{schab2018lower,kundu2016information}:
\begin{equation}
\sub{\C}=\U^{\mathrm{H}} \C \U
\end{equation}
where $\U=\R^{\frac{1}{2}} \M(\M^{\mathrm{H}} \R \M)^{-\frac{1}{2}}$ is a unitary matrix with $\U^H\U=\mathbf{I}_{\bar{K}}$, the identity matrix of rank ${\bar{K}}$.

Arranging the eigenvalues of $\C$ and $\sub{\C}$ in descending order, and invoking the Poincar\'{e} Separation Theorem \cite{Horn2012} (see Appendix B), it can be shown that the eigenvalues of $\C$ bound those of $\sub{\C}$ as,  
\begin{equation}
\label{eq:PSTResult1}
\lambda_k  \geqslant \bar{\lambda}_k \geqslant \lambda_{k+K-\bar{K}}, 1\geqslant k\geqslant \bar{K}.
\end{equation} 
for all frequencies, where $K$ is the rank of the super-structure Z matrix, and $\sub{K}$ is the rank of the substructure matrices.

\section{Poincar\'{e} Separation Theorem}
\label{Appendix_A}
Let $M_K$ be the space of all square matrices of dimension $K$, and $\C\in M_K$ be Hermitian.  Suppose that $1\leq \bar{K}\leq K$, and let $U_1$, $\cdots$, $U_{\bar{K}}$ be orthogonal, namely $\U^{\mathrm{H}}\U=\mathbf{I}_{\bar{K}}$. Let the eigenvalues of $\C$ and $\U^{\mathrm{H}}\C\U$ be arranged in algebraic descending order. Then \cite{Horn2012},

\begin{equation}
\lambda_i (\C)\geq \lambda_i (\U^{\mathrm{H}}\C\U)\geq \lambda_{i+K-\bar{K}} (\C), i=1,\cdots,\bar{K}
\end{equation}

\noindent and when arranging the eigenvalues of $\C$ and $\U^{\mathrm{H}}\C\U$ in algebraic ascending order, then

\begin{equation}
\lambda_i (\C)\leq \lambda_i (\U^{\mathrm{H}}\C\U)\leq \lambda_{i+K-\bar{K}} (\C), i=1,\cdots,\bar{K}.
\end{equation}

\bibliographystyle{IEEEtran}
\bibliography{main}

\end{document}